\documentclass[12pt]{article}

\usepackage{amsmath,amssymb}
\usepackage{a4wide}

\makeatletter
\renewcommand\section{\@startsection {section}{1}{\z@}%
                                   {-3.5ex \@plus -1ex \@minus -.2ex}%
                                   {2.3ex \@plus.2ex}%
                                   {\normalfont\large\bfseries}}
\renewcommand\subsection{\@startsection{subsection}{2}{\z@}%
                                     {-3.25ex\@plus -1ex \@minus -.2ex}%
                                     {1.5ex \@plus .2ex}%
                                     {\normalfont\normalsize\bfseries}}

\makeatother

\newtheorem{thm}{Theorem}[section]
\newtheorem{lem}[thm]{Lemma}
\newtheorem{cor}[thm]{Corollary}

\newtheorem{defi}[thm]{Definition}
\newtheorem{rem}[thm]{Remark}
\newtheorem{pty}[thm]{Property}

\def\K{{\sf K}}

\def\NN{{\mathbb{N}}}

\def\diag{{\rm diag}}
\def\rank{{\rm rank}}

\def\MM{{\sf MM}}
\def\so{{O {\;\!\tilde{}}\,}}
\newenvironment{mat}{\left[\!\begin{array}}{\end{array}\!\right]}

\title{
{\normalsize \bfseries ASYMPTOTICALLY FAST POLYNOMIAL MATRIX 
\\[-0.2cm] ALGORITHMS FOR MULTIVARIABLE SYSTEMS }}

\begin{document}

\date{}


\author{{\normalsize
Claude-Pierre Jeannerod
 and Gilles Villard}\\
{\footnotesize \em CNRS, INRIA, Laboratoire LIP, \'Ecole normale sup\'erieure de Lyon} \\[-0.2cm]
{\footnotesize \em 46, All\'ee d'Italie, 69364 Lyon Cedex 07, France}\\
}

\maketitle

\def\thefootnote{}
\footnotetext[1]{\hspace*{-1cm}
\begin{tabular}{ll}URLs: & {\tt http://perso.ens-lyon.fr/claude-pierre.jeannerod}\\
& {\tt http://perso.ens-lyon.fr/gilles.villard}\end{tabular}\\

\newcount\hour\newcount\minutes\newcount\temp
\hour=\time \divide\hour by 60 \temp=\hour \multiply\temp by 60
\minutes=\time \advance\minutes by-\temp

{\hfill {\em International Journal of Control Submission ---
    \the\day/\the\month/\the\year} }}
\def\thefootnote{\fnsymbol{footnote}}

\begin{abstract}
We present 
the asymptotically fastest known algorithms
for some basic problems on univariate polynomial matrices:
{\em rank, nullspace, determinant, 
generic 
inverse, reduced
  form}~\cite{GJV03-2,JeVi04,Sto03,StVi05-2}.
We show that they essentially can be reduced to two computer algebra techniques,
{\em minimal basis computations} and {\em matrix fraction expansion/reconstruction},
and to polynomial matrix multiplication.
Such reductions eventually imply that all these problems can be solved
in about the same amount of time as polynomial matrix multiplication.
\end{abstract}


\section{Introduction} \label{sec:intro}

We aim at drawing attention to today's 
asymptotically fastest known algorithms for computing with polynomial matrices.
In particular, we shall focus on the following problems:
compute the {\em rank}, a right or {\em left nullspace}, the {\em determinant}, the {\em inverse} 
and a column- or {\em row-reduced form} of a given polynomial matrix.
Polynomial matrices are quite common in the analysis 
of multivariable linear systems and Kailath's treatise 
{\em Linear Systems}~\cite{Kai80} is a good illustration of this.

Recently, algorithms have been
designed~\cite{GJV03-2,JeVi04,Sto03,StVi05-2} that 
allow to compute solutions to these problems 
in essentially the same amount of time as when multiplying two polynomial matrices together.
More precisely, given a field $\K$---for example the complex numbers, the rationals or a finite
field---and given a polynomial matrix $A \in \K[x]^{n \times n}$ whose entries have degree in $x$ bounded by $d$,
these algorithms allow to compute
$\rank\, A$, $\ker A$, $\det A$ and to row-reduce $A$ in 
$\so(n^\omega d)$ operations in $\K$, and to compute $A^{-1}$ when $A$
is generic 
in $\so(n^3 d)$ operations in $\K$.
Here, $\so(n^\omega d)$ is the best known asymptotic bound for multiplying
two matrices in $\K[x]^{n \times n}$ of degree
$d$~\cite{CaKa91,BoSc05}, 
where $2 \leq \omega < 2.376$ is the exponent of 
matrix multiplication over $\K$~\cite[Chapter 15]{BCS97}. 
Using schoolbook matrix multiplication, we have $\omega = 3$ and the bound $\so(n^\omega d)$ becomes $\so(n^3 d)$.
Furthermore, the soft-O notation $\so$ simply indicates some missing logarithmic factors of the form 
$\alpha {(\log n)}^\beta {(\log n)}^\gamma$ for three positive real numbers $\alpha, \beta, \gamma$.
By achieving the complexity estimate $\so(n^\omega d)$, these 
algorithms improve upon all the complexity estimates that were known previously.

In this paper, evidence is given that the key tools for such improvements are:
\begin{itemize}
\item Minimal bases of $\K[x]$-modules;
\item Expansion/reconstruction of polynomial matrix fractions.
\end{itemize}
The former has the same flavour as in~\cite{For75} while for the
fractions  we heavily 
rely on the concepts in~\cite[Chapter 6]{Kai80}.
Two kinds of minimal bases, namely {\em approximant bases} and {\em nullspace bases},
are studied in Section~\ref{sec:approx}. There we will see that
such bases are small enough to be computed fast, that is, in $\so(n^\omega d)$
operations in $\K$.
Polynomial matrix fractions are matrices $F \in \K(x)^{n \times n}$, where $\K(x)$
is the field of rational functions over $\K$. By expansion of $F$, 
we thus mean a power series expansion $F = \sum_{i=0}^\infty F_i x^i \in \K[[x]]^{n \times n}$ and by reconstruction
of $F$ we mean a left or right quotient of polynomial matrices like $F = A^{-1} B$ or $F = B A^{-1}$.
It turns out that all we need is truncated expansions and reconstructed quotients
of rather low degree, both of which can be computed fast as seen in Section~\ref{sec:fractions}.
The key idea here is that an {\em approximant} of sufficiently high
order---with respect to the input problem---may lead to an {\em exact} solution over $\K[x]$.
This is well-known in computer algebra, at least for scalar rational functions~\cite[\S5.7]{vzGG99},
but as far we know the extension to the matrix case is more 
recent~\cite{GJV03-2,JeVi04,Sto03,StVi05-2}.

Minimal bases and matrix fractions are interesting not only because
they can be computed fast, but also---and, perhaps, mainly---because 
computing a minimal basis and expanding/reconstructing a matrix fraction 
are problems to which we can reduce all other problems like {\em rank}, 
{\em left nullspace}, {\em determinant}, {\em generic 
inverse} and {\em row-reduced form}.
The goal of Section~\ref{sec:applis} is precisely to show this:
there the above problems are thus seen as applications of the techniques studied in
Sections~\ref{sec:approx} and~\ref{sec:fractions}.

If we assume given an $\so(n^\omega d)$ algorithm for multiplying two $n$ by $n$
polynomial matrices of degree $d$, combining the reductions of Section~\ref{sec:applis}
with the cost estimates of Sections~\ref{sec:approx} and~\ref{sec:fractions} then yields
$\so(n^\omega d)$ solutions to all our problems under consideration.
Of course, we could have introduced a cost function $\MM(n,d)$ for polynomial
matrix multiplication and derived more precise complexity estimates for each
of the problems, in terms of (functions of) $\MM(n,d)$. 
However, we prefer for this paper to stick to the more readable $\so(n^\omega d)$ bound,
which already gives a good sense of the link with polynomial matrix multiplication.



A first task remaining would be to relax the regularity assumptions made
for {\em inversion} (the input should be generic and of 
dimensions a power of two, see Section~\ref{subsec:appli_inv}) 
and for {\em row-reduction} (the input should be non-singular, see Section~\ref{subsec:appli_rowred}).
But even these ``generic'' situations are enough for our purpose here of showing how
to rely on minimal bases and matrix fraction expansions/reconstructions.

Also, recently, other problems on polynomial matrices than those treated in this paper have been
shown to have about the same complexity as polynomial matrix multiplication. 
An example is the problem of computing the {\em Smith normal form} and thus also
the determinant, 
whose solution in~\cite{Sto03} gives us Theorem~\ref{theo:expansion}. 
However---and this is the second task remaining---, the 
list of problems that can be solved in
about the same number of operations as for
polynomial matrix multiplication still has to be augmented.
The question is particularly interesting for the problem of computing the 
{\em characteristic polynomial} and the {\em Frobenius normal form},
for which the best known solutions~\cite{Kal92,KaVi04-2} 
have cost $\so (n^{2.7}d)$ still greater than $\so (n^{\omega}d)$.
\\\\{\em Notation and basic reminders.}
Here and hereafter $\log$ denotes the logarithm in base two
and $I_n$ the $n$ by $n$ identity matrix. 
For a matrix $A$ over $\K[x]$, we denote its value at $x=0$ by $A(0)$.
For  $d \in \NN$ and a matrix $F$ over $\K[[x]]$, $F \equiv 0 \mod x^d$ 
means that each entry of $F$ is a multiple of $x^d$,
and $F \mod x^d$ means that we truncate $F$ into a polynomial matrix where
only powers in $x$ strictly less than $d$ appear.
By {\em size} of a polynomial matrix over $\K[x]$ we mean the number of
elements of $\K$ that are necessary to represent it.
For example, $M \in \K[x]^{n \times m}$ of degree $d$ has size at most $nm(d+1) = O(nmd)$.
A polynomial matrix is said to be {\em non-singular} when it is square
and when its determinant is a non identically zero polynomial.
Two matrices $A, R \in \K[x]^{n \times n}$ are {\em unimodularly left equivalent}
when there exists $U \in \K[x]^{n \times n}$ such that $\det U$ is a non-zero
constant---that is, $U$ is {\em unimodular}--- and when $UA = R$.


\section{Minimal approximant bases and minimal nullspace bases} \label{sec:approx}

Our solutions for solving a class of polynomial matrix problems
in about the
same number of operations in $\K$ as for multiplying two polynomial
matrices will fundamentally rely on computing 
minimal bases of $\K[x]$-modules. The target complexity estimate
$\so (n^{\omega}d)$ 
is reached since the bases we use are small, with size $O(n^2d)$ is
most cases, and may be computed fast (see Theorem~\ref{theo:costapprox}
below).

\begin{defi} \label{def:minbas} 
Let $\mathcal M$ be a $\K[x]$-submodule of $\K[x]^n$ of dimension $\mathcal D$.
A basis $N_1, \ldots , N_{\mathcal D} \in \K[x]^{n}$ of $\mathcal M$ with degrees $\delta _1
\leq \cdots \leq \delta _{\mathcal D}$ is called a {\em minimal basis} 
if any other basis of $\mathcal M$ with degrees $d_1\leq \cdots \leq d _{\mathcal D}$ 
satisfies $d  _i \geq \delta _i$ for $1 \leq i \leq \mathcal D$.
The degrees $\delta_i$ are called the {\em minimal indices} of $\mathcal M$.
\end{defi}

In applications to multivariable systems, this definition follows the
study of minimal polynomial bases of vector spaces in~\cite{For75}.
The two important examples of such bases that we use in this paper 
are minimal approximant bases and minimal nullspace bases. The
approximant bases are defined from a power series matrix $F$ over
$\K[[x]]$, the nullspace bases are computed as special approximant
bases from a polynomial matrix $F=A$ over $\K[x]$.

\subsection{Minimal approximant bases} \label{subsec:mab}
Given a formal power series $F \in \K[[x]]^{n \times n}$ and an order $d \in \NN$, 
we take for $\mathcal M$ the set of all approximants for $F$ of order $d$:
\[ 
\mathcal M = \{ v \in \K[x]^{1 \times n} \,\,:\,\, v F \equiv 0 \mod x^d\}.
\]
The minimal bases of $\mathcal M$ 
are called {\it minimal approximant bases for $F$
of order $d$}. Since $\mathcal M$ has dimension $n$, such bases 
form non-singular $n \times n$ polynomial matrices. 
These polynomial matrices further have degree up to $d$ and their
size is thus of the order of $n^2 d$.
\begin{thm}~{\em \cite{GJV03-2}.} \label{theo:costapprox}
Let $F \in \K[[x]]^{n \times n}$ and $d \in \NN$.
A minimal approximant basis 
for $F$ of order $d$ 
can be computed
in $\so(n^\omega d)$ operations in $\K$.
\end{thm}

Our notion of minimal approximant bases is directly inspired
by~\cite{BeLa94} with some adaptations for fully reflecting
the polynomial matrix point of view.
The cost estimate of Theorem~\ref{theo:costapprox} is a
matrix polynomial generalization  of the 
recursive
Knuth/\allowbreak Sch\"on\-hage half-gcd algorithm for scalar
polynomials~\cite{Knu70,Sch71}
(see also~\cite[\S11.1]{vzGG99}), that takes into account fast
polynomial matrix multiplication.

For a matrix $A$ over $\K[x]$,
we denote by $d_i$ its $i$th {\it row degree}, that is,
the highest degree of all the entries of the $i$th row of $A$.
The {\it row leading matrix} of $A$ is the constant matrix whose $i$th row
consists of the coefficients of $x^{d_i}$ in the $i$th row of $A$.
We recall from~\cite[\S6.3.2]{Kai80} that a full row rank $A$ is
{\it row-reduced} when its
row leading matrix also has full rank.
As a consequence of their minimality, minimal approximant bases have the
following properties, which will be used in Section~\ref{subsec:minnullspace} 
when specializing approximants for power series matrices 
to approximants for polynomial matrices.
\begin{pty} \label{pty:pty}
Let $N$ be a minimal approximant basis for $F$ of order $d$.
Then,\\[-0.75cm]
\begin{enumerate}
\item[{\sc i.}] $N$ is row-reduced;\\[-0.75cm]
\item[{\sc ii.}] If $v \in \mathcal M$ has degree at most $d$, then 
there is a unique $u \in \K[x]^{1 \times n}$ such that $v = u N$.
Furthermore, $N$ has at least one row of degree at most $d$.
\end{enumerate}
\end{pty}

Property~{\sc i} above is a consequence of the minimality of the
basis~\cite[Theorem~6.5-10]{Kai80}. Property~{\sc ii} is the fact
that the rows of $N$ form a basis, together with the predictable
degree property~\cite[Theorem~6.3-13]{Kai80}.

\subsection{Minimal nullspace bases} \label{subsec:minnullspace}
Given a polynomial matrix $A \in\K[x]^{n \times n}$ of rank $r$, we now take
\[ 
\mathcal M = \{ v \in \K[x]^{1 \times n} \,\,:\,\, v A = 0\}.
\]
This is a $\K[x]$-submodule of $\K[x]^n$ of dimension $n-r$.
Its bases are called {\it minimal nullspace bases for $A$}
and form full rank $(n-r) \times n$ polynomial matrices.
The minimal indices $\delta _1 \leq \cdots\leq \delta _{n-r}$ (see
Definition~\ref{def:minbas}) are  called 
the (left) {\em Kronecker indices} of $A$~\cite[\S6.5.4]{Kai80}. 
For any given degree threshold $\delta$, we further define
\[
\kappa = \max \{1\leq i \leq n-r \,\,:\,\, \delta _i \leq \delta\}.
\]
A corresponding family of $\kappa$ linearly independent vectors of degrees $\delta _1,
\cdots, \delta _{\kappa}$ is a family of {\em minimal nullspace vectors} of degree at
most $\delta$.
The theorem below says that if $F = A$ is a polynomial matrix then
any minimal approximant basis for $A$ of sufficiently high
order actually contains a family of minimal nullspace vectors for $A$.
\begin{thm} \label{theo:approxnull}
Let $A \in \K[x]^{n \times n}$ be of degree $d$.
Let $N$ be a minimal approximant basis for $A$ of order $\delta + d + 1$. 
Then exactly $\kappa$ rows of $N$ have degree at most $\delta$;
these rows 
are in the (left) nullspace of $A$ 
and their degrees are the Kronecker indices $\delta _1,\ldots ,\delta _{\kappa}$. 
\end{thm}
\begin{proof}
A row $v$ of $N$ of degree bounded by $\delta$ satisfies 
$v A \equiv 0 \mod x^{\delta + d + 1}$, and 
using $\deg v A \leq \delta +d$, 
$vA=0$. Let $k$ be the number of such $v's$ in $N$, 
from the definition of $\kappa$ and since $N$ is non-singular,  
$k \leq \kappa$. 
We now verify that $k \geq \kappa$. We consider $\kappa$ linearly
independent vectors
$v_i$ of degrees
$\delta _i$ in the nullspace of $A$. From Property~\ref{pty:pty} we 
have 
$v_1 = u_1 M$ and deduce that one row of
$N$ has degree bounded by $\delta _1$. Now, if $N$ has $i-1$ rows of
degrees bounded by $\delta _1, \ldots , \delta _{i-1}$, then the
same reasoning with $v_i$ as for $v_1$ shows that $N$ has a row of
degree bounded by $\delta _i$, linearly independent with respect to
the first $i-1$ chosen ones. It follows that $k \geq \kappa$ 
rows of $N$ have degrees bounded by $\delta _1, \ldots , \delta
_{\kappa}$, and are in the nullspace of $A$. Hence $k=\kappa$, and
we conclude using 
Definition~\ref{def:minbas} and 
the
minimality of the $\delta _i$'s.
\end{proof}
For some applications, a {\em shifted degree} may be introduced
(see~\cite{BLV04}
and the references therein), and some aspects of Theorem~\ref{theo:approxnull}
may be generalized accordingly (see \cite[Theorem~4.2]{BLV04} or
\cite[Lemma~6.3]{StVi05-2}). 

Notice that if the Kronecker indices of $A$ are all bounded by $d$
then an {\em entire} minimal nullspace basis for $A$ can already be computed fast:
by Theorem~\ref{theo:approxnull}, it suffices to compute a minimal 
approximant basis for $A$ of order $2d+1$ and, by Theorem~\ref{theo:costapprox},
this computation can be done in time $\so(n^\omega d)$.

However, in the general case of unbalanced degrees, 
computing a nullspace basis fast is much less immediate
and the method we shall give in Section~\ref{subsec:appli_nullspace} 
relies on the complexity result given below
in Theorem~\ref{theo:partialnullspace}.
The cost given here is the one of a randomized
algorithm of the Las Vegas kind---always correct, probably
fast. 
The algorithm
outputs correct minimal vectors
in time $\so{(n^{\omega}d)}$ with good probability, say greater than
$1/2$,
otherwise
returns {\em failure} (a correct result will be obtained after repetition).

\begin{thm}~{\em \cite{StVi05-2}.} \label{theo:partialnullspace}
Let $A \in \K[x]^{(n+m) \times n}$ with $m \leq n$ 
be of full column rank and degree bounded by $d$.
If $\delta \in \NN$ satisfies
\begin{equation} \label{eq:compro2}
{\delta m} = O(nd), 
\end{equation}
then a family of minimal nullspace vectors of degree at most $\delta$
can be computed by a randomized Las Vegas (certified) algorithm
in $\so(n^\omega d)$ operations in $\K$.
\end{thm}
Note that the cost estimate $\so (n^{\omega}d)$ relies
on the compromise~(\ref{eq:compro2}) between the minimal nullspace vector degree bound $\delta$ 
and the row dimension of matrix $A$. 
For example, when $m=1$ one can compute a nullspace vector of degree as large as $O(nd)$,
whereas when $m=n$ one may compute up to $n$ nullspace vectors of degree $O(d)$.
Random values are introduced essentially through a random compression
matrix $P \in \K[x] ^{n \times m}$ that allows to compute minimal
vectors more efficiently using the matrix $AP \in \K[x] ^{(n+m)
  \times m}$ rather than directly from $A \in \K[x] ^{(n+m)
  \times m}$ (see \cite[Proposition~5.4]{StVi05-2}).


\section{Matrix fraction expansion and reconstruction} \label{sec:fractions}

{\em Matrix fraction expansion} and {\em reconstruction} will be key
tools especially for the row reduction and the nullspace problems.
Fraction reconstruction is a useful tool in computer algebra 
({\em e.g.} see~\cite[\S5.7]{vzGG99} for scalar polynomials), that
is directly connected to {\em coprime factorization} (see below, and
\cite[Chapter~6]{Kai80} or \cite{OaVa99} and the references therein).

For a polynomial matrix $A$ that is non-singular at $x=0$ and a polynomial matrix $B$, 
the techniques of~\cite[Proposition~17]{Sto03} 
reduce the computation of parts of the power series expansion 
$$A^{-1}B = \sum_{i = 0}^\infty F_i x^i$$ to polynomial matrix multiplication. 
By parts of the expansion, we mean a given number of consecutive matrix coefficients $F_i$. 
This is summarized in the following theorem.
\begin{thm} {\em \cite{Sto03}.}  \label{theo:expansion}
Let $A \in \K[x] ^{n \times n}$ with $A(0)$ non-singular, and
$B \in \K[x] ^{n \times m}$. Assume that $A$ and $B$ have degree
bounded by $d$ and let $h \in \NN$ be such that $h = O(nd)$. 
If $\delta \in \NN$ satisfies
\begin{equation} \label{eq:compro}
{\delta m} = O(nd), 
\end{equation}
then the $\delta$ coefficients $F_{h}, F_{h+1},\ldots,F_{h+\delta-1} \in \K^{n \times m}$ 
of the expansion of $A ^{-1}B$ at $x=0$ can
be computed  in $\so (n^{\omega}d)$ operations in~$\K$.
\end{thm}

Similarly to Theorem~\ref{theo:partialnullspace}, the cost estimate $\so (n^{\omega}d)$ relies
on the compromise~(\ref{eq:compro}) between approximation
order $\delta$ and the column dimension of matrix $B$. 
For instance, for a vector $B=b \in \K[x]^{n \times 1}$ and $h = 0$, 
one can expand $A ^{-1}b$ up to order $O(nd)$, whereas 
with $B=I_n$ and $h = 0$, one gets the expansion of $A ^{-1}$ up to order $O(d)$. 
In Section~\ref{subsec:appli_rowred}, we shall use this result with 
$B=I_n$ and $h = (n-1)d+1$ in order to get a high-order slice of length $O(nd)$ 
of the expansion of $A^{-1}$.

Notice also that the regularity assumption $\det A(0) \neq 0$ in
Theorem~\ref{theo:expansion} is not restrictive. Indeed, it can 
be satisfied with high probability using random shifts, thus
yielding randomized algorithms for any $A(0)$.
Typically, with a randomly chosen $x_0 \in \K$, 
we shift $x$ in the input like $x \gets x+x_0$
to get a regular input at zero and, at the end of the computation, 
we shift $x$ back like $x \gets x-x_0$ to recover the result 
(see~\cite{Sto03,GJV03-2,StVi05-2}).\\

A rational matrix $H \in \K(x)^{n \times m}$ 
is {\em strictly proper} if $\lim _{x\to  \infty} H(x) = 0 \in \K^{n \times m}$.
In most applications, difficulties arise when $A ^{-1} \in \K(x)^{n \times n}$ is {\em not}
strictly proper. However, one can define another 
fraction that is always strictly proper and shares  
some invariants with $A ^{-1}$. Before seeing this, 
we first need to recall some facts about {\em greatest common divisors}
of two polynomial matrices.

\begin{defi}\label{def:gcd}
A (left) {\em matrix gcd} of $A \in \K[x] ^{n \times n}$ and $B \in \K[x]
^{n \times m}$ is any full column rank polynomial matrix 
$G$ such that 
$[G~~0] U = [A ~~ B]$
with $U$ unimodular over~$\K[x]$. 
\end{defi}

Definition~\ref{def:gcd} is for instance from~\cite[Lemma 6.3-3]{Kai80}. 
If $[A ~~ B]$ has full row rank then all the gcd's of $A$ and $B$ are 
non-singular and equivalent 
with respect to multiplication on the right 
by any unimodular matrix in $\K[x] ^{n \times n}$ (see
\cite[Lemma 6.3-4]{Kai80}). 
A non-singular $A \in \K[x] ^{n \times n}$ is said to be (left) {\em coprime} with 
$B \in \K[x]^{n \times m}$ if any gcd of $A$ and $B$ is unimodular 
({\em the} gcd may be chosen as being the identity matrix $I_n$).
Similar definitions hold for rights gcd's and right coprimeness.

\begin{thm}~{\em \cite{GJV03-2}.} \label{theo:expproper}
Let $A \in \K[x] ^{n \times n}$ of degree bounded by $d$, with $A(0)$ non-singular.
For $A^{-1} = \sum_{i=0}^\infty F_i x^i$ and 
$h > (n-1)d$, let 
$H \in \K(x)^{n \times n}$ be given by 
$H =\sum_{i=0}^\infty F_{h+i} x^i$.
Then $H = A^{-1} (AH) = (HA) A^{-1}$ is strictly proper, and
$AH$ and $HA$ are polynomial matrices that are respectively left and right 
coprime with $A$.
\end{thm}
\begin{proof}
Let $B = AH$. 
By definition of $H$ we have 
$I_n = A (A ^{-1} \mod x ^h) + x^h B$
which in~\cite{Sto03} is (17) on the left with 
$B$ and $T$ respectively set to $I_n$ and $A$.
It follows that $B$ is a polynomial matrix.
On the other hand, $H = A ^{-1}B$ is strictly proper because 
$A ^{-1}B = x ^{-h}A ^{-1} - x ^{-h}  (A ^{-1} \mod x ^h)$
and $h > (n-1)d \geq \deg A ^{*}$ where $A ^{*}$ is the adjoint
matrix of $A$. For establishing coprimeness we use  
\begin{equation}\label{eqdem1}
[A ~~ x^h B] \left[ \begin{array}{cc}
I_n & (A ^{-1} \mod x ^h) \\ 0 & I_n 
\end{array}\right] 
\left[
\begin{array}{cc}
0 & I_n  \\ I_n & -A  \end{array}\right] 
= [I_n ~~ 0],
\end{equation}
and the fact that if $G$ is a left gcd of $A$ and $B$ it satisfies
\begin{equation}\label{eqdem2}
[G~~0]U = [A~~B]
\end{equation}
with $U$ unimodular.
Identities~(\ref{eqdem1}) and~(\ref{eqdem2}) give that their exists
a polynomial 
matrix $V$ such that $[G~~0]V=[I_n ~~0]$, hence a polynomial matrix
$W$ such that $GW=I_n$.  Since $G$ is a polynomial matrix this implies
that $G$ is unimodular, and $A$ and $B$ are left coprime. 
With $B = HA$, one could show similary right coprimeness.
\end{proof}

For our application in Section~\ref{subsec:appli_rowred},
we will need only the first, say $\delta$, coefficients of the expansion of 
$H$ as in Theorem~\ref{theo:expproper}.
These coefficients thus correspond to a slice of order $h$
and length $\delta$ of the expansion of $A^{-1}$ and, 
to recover them, we shall use Theorem~\ref{theo:expansion} with $B=I_n$.

Matrix power series expansion will be used in conjunction with 
matrix {\em (irreducible) fraction reconstruction} or, equivalently, 
{\em (coprime) factorization}.
We show below that minimal approximant bases are appropriate tools 
for solving  these problems.

\begin{defi}
A (left) {\em factorization of degree $\delta$} of  
a rational matrix $H \in \K(x)^{n \times n}$ is a representation 
$H=V ^{-1}U$ with $U$ and $V$ two polynomial matrices of degree
bounded by $\delta$.
This factorization is said to be {\em coprime} when $U$ and $V$ are (left) coprime.
\end{defi}

A similar definition holds on the right.
Hence, given $H \in \K(x)^{n \times n}$, the reconstruction or factorization problem is to 
recover $U$ and $V$ over $\K[x]$ such that $V ^{-1} U=H$.
If $H$ is defined at $x=0$ and given by its formal expansion $F \in \K[[x]]^{n \times n}$, this  
problem reduces to computing a suitable $[U~~V] \in \K[x]^{n \times 2n}$ such that 
$$
\begin{mat}{cc} U & V \end{mat} \!\! \begin{mat}{c} -I_n \\ F \end{mat} = 0.
$$

\begin{thm} \label{theo:recon}
Let $H\in \K(x)^{n \times n}$ be strictly proper, 
with expansion $F\in \K[[x]]^{n \times n}$ at $x=0$. 
Assume that $H$ admits a right factorization
of degree $\delta _R$ and a left factorization 
of degree $\delta _L$. 
Let $N \in \K[x] ^{2n \times 2n}$ be a minimal approximant basis 
for $[-I_n~~F ^T]^T$ of order $\delta _L + \delta _R +1$. 
Then exactly $n$ rows of $N$ have degree bounded by $\delta _L$;
these rows form a matrix $[U~~V] \in \K[x]^{n \times 2n}$ such that $V^{-1}U$ is a left coprime
factorization of $H$, with $V$ row-reduced. 
\end{thm}
\begin{proof}
Let $B A ^{-1}$ be a right factorization of $H$ of degree $\delta_R$ 
and $T ^{-1}S$ be a left factorization  of $H$ degree $\delta _L$.
Since $[-I_n~~F ^T]^T A = [-A ^T ~~ B ^T]^T$, $N$ is also a
minimal approximant basis of the latter matrix whose rank is $n$.
Using $[S~~T] [-A ^T ~~ B ^T]^T=0$, with the threshold $\delta = \delta _L$ 
we have $\kappa = n$. 
(See before Theorem~\ref{theo:approxnull} for a definition of $\kappa$.)
Hence, applying Theorem~\ref{theo:approxnull} to $[-A ^T ~~ B ^T]^T$ 
(augmented on the right with $n$ zero columns) 
with $\delta = \delta _L$ and $d=\delta _R$, we know that 
exactly $n$ rows of $N$ have degree bounded by $\delta _L$ 
and are in the nullspace of  $[-A ^T ~~ B ^T]^T$. 
We denote the corresponding matrix by $[U~~V]$. 
The matrix $V$ is non-singular, for otherwise there would be a non-zero
vector $v$ such that $vV=0$. This would imply $vVB=vUA=0$, hence either
$vU=0$ or $wA=0$ for $w=vU\neq 0$, and would contradict either that 
$\text{rank}[U~~V]=n$ or that $A$ is non-singular.
Therefore, $V ^{-1}U$ is a left factorization of $H$. 

This factorization must further be left coprime.
Indeed, non-coprimeness would imply that $U$ and $V$ have a non-trivial left gcd, 
that is, there exists a polynomial matrix $G$ such that $U=GU'$, $V=GU'$ 
and $\deg (\det G) > 0$. 
Then $[GU'~~GV']$ would be a submatrix of the
minimal approximant basis, which would contradict its irreducibility
in~\cite[Theorem 6.5-10]{Kai80} by considering a zero of $\det G$.
In addition, the fact that $[U~~V]$ as a submatrix of $N$ is
row-reduced (see Property~\ref{pty:pty}), implies that $V$ is row-reduced. 
Indeed, since $H=V^{-1}U$ is strictly proper, the row degrees of $U$ are
strictly smaller than those of $V$~\cite[Lemma 6.3-10]{Kai80}, 
and the row leading matrix of $[U~~V]$ has the form $[0~~L]$
where $L$ is the row leading matrix of $V$, which is
then non-singular.
\end{proof}

As an immediate consequence of Theorem~\ref{theo:recon} and 
Theorem~\ref{theo:costapprox}, coprime factorizations can be computed 
fast when the input matrix fractions admit left and right 
factorizations of degree $O(d)$. This corollary, given below, 
will be applied in Section~\ref{subsec:appli_rowred} 
to the particular matrix fraction of Theorem~\ref{theo:expproper}.
\begin{cor} \label{cor:trivial_recon}
Let $H \in \K(x)^{n \times n}$ be as in Theorem~\ref{theo:recon} with $\delta _L = O(d)$ and $\delta_R = O(d)$.
Given the first $\delta_L+\delta_R+1$ coefficients of the expansion of $H$ at $x=0$, 
one can compute a left coprime factorization of $H$ in $\so(n^\omega d)$ operations in $\K$.
\end{cor}




\section{Applications} \label{sec:applis}
In this section, we show how the techniques presented in
Sections~\ref{sec:approx} and~\ref{sec:fractions} 
can be used to solve the following problems asymptotically fast:
\begin{itemize}
\item ${\sf Inv}_{n,d}$: given a non-singular $A \in \K[x]^{n \times n}$ of degree $d$, compute $A^{-1}$.
\item ${\sf Det}_{n,d}$: given $A \in \K[x]^{n \times n}$ of degree $d$, compute $\det A$.
\item ${\sf RowRed}_{n,d}$: given $A \in \K[x]^{n \times n}$ of degree $d$, compute a row-reduced form of $A$.
\item ${\sf Nullspace}_{n,d}$: given $A \in \K[x]^{n \times n}$ of degree $d$,
compute the rank $r$ of $A$ and a full rank $N \in \K[x]^{(n-r) \times n}$ such that $N A = 0$.
\item ${\sf Factor}_{n,d}$: given a right factorization of degree $d$ of $H \in \K(x)^{n \times n}$, 
compute a left factorization of $H$. 
\end{itemize}
Our approach here is to reduce each of the above five problems to (collections of)
the problems below, for which $\so(n^\omega d)$ solutions are known:
\begin{itemize}
\item ${\sf MatMul}_{n,d}$: 
given $A, B \in \K[x]^{n \times n}$ of degree $d$, compute the product $AB$.

$\hookrightarrow$ {\em for solutions in time $\so(n^\omega d)$ see~\cite{CaKa91}, \cite{BoSc05}.}

\item ${\sf PartialNullSpace}_{m,\delta}$: 
given $\delta =  O(nd/m)$ with $n,d$ fixed, and given $A \in \K[x]^{(n+m) \times n}$
of degree $d$, compute the minimal nullspace vectors of $A$ of degree at most $\delta$.

$\hookrightarrow$ {\em solved in time $\so(n^\omega d)$ by Theorem~\ref{theo:partialnullspace}.}

\item ${\sf MatFracExp}_{m,\delta}$: 
given $\delta =  O(nd/m)$ with $n,d,h$ fixed such that $h = O(nd)$, and given $A \in \K[x]^{n \times n}, B \in \K[x]^{n \times m}$ 
of degree $d$ with $A(0)$ non-singular, 
compute the $\delta$ coefficients $F_{h}, F_{h+1},\ldots,F_{h+\delta-1}$ 
of the expansion of $A^{-1} B$ at $x=0$.

$\hookrightarrow$ {\em solved in time $\so(n^\omega d)$ by Theorem~\ref{theo:expansion}.}

\item ${\sf MatFracRec}_{n,d}$: 
given $\delta_L, \delta_R = O(d)$ and the first $\delta_L+\delta_R+1$ coefficients of the expansion at $x=0$ 
of $H \in \K(x)^{n \times n}$ as in Theorem~\ref{theo:recon}, 
compute a left coprime factorization of $H$ with row-reduced denominator.

$\hookrightarrow$ {\em solved in time $\so(n^\omega d)$ by Corollary~\ref{cor:trivial_recon}.}
\end{itemize}
Assuming that $n$ is a power of two and given a problem ${\sf P}_{n,d}$ or ${\sf P}_{m,\delta}$ 
such as any of those just introduced, we define the collections of problems we shall rely on as
\begin{equation} \label{eq:pstar}
{\sf P}_{n,d}^* := \left\{\mbox{solve $O(2^i)$ problems ${\sf P}_{n/2^i, 2^i d}$}\right\}_{0 \leq i < \log n}.
\end{equation}
Such collections can be solved at about the same cost as polynomial matrix multiplication,
as shown below. Here subscripts $n,d$ and $m,\delta$ should be added to ${\sf P}$ and ${\sf P}^*$
depending on the underlying problem.
\begin{lem} \label{lem:costpstar}
For all ${\sf P} \in \{ {\sf MatMul},{\sf PartialNullSpace}, {\sf MatFracExp}, {\sf MatFracRec} \}$,
one can solve ${\sf P}^{*}$ in $\so(n^\omega d)$ operations in $\K$.
\end{lem}
\begin{proof} This an immediate consequence of (\ref{eq:pstar}) and of the bound $\so(n^\omega d)$ on 
the cost of each of these four problems.
\end{proof}

\subsection{Polynomial matrix inversion (${\sf Inv}_{n,d}$)} \label{subsec:appli_inv}

{\em Given $A \in \K[x]^{n \times n}$ non-singular of degree $d$, the problem is to compute $A^{-1} \in \K(x)^{n \times n}$.}\\

Assuming that $A$ is generic and that $n$ is a power of two,
we recall from~\cite{JeVi04} how ${\sf Inv}_{n,d}$ reduces to
${\sf PartialNullSpace}_{n,d}^*$ plus some polynomial matrix multiplications.
The algorithm in~\cite[p.75]{JeVi04} essentially consists in computing in $\log n$ steps
a  non-singular matrix $U\in\K[x]^{n\times  n}$ and a diagonal matrix $B\in\K[x]^{n\times n}$ such that
\begin{equation} \label{eq:UA=B} 
UA=B.
\end{equation} 
The  inverse of $A$   is then recovered as $A^{-1}  = B^{-1}U$.
The first step is as follows. 
Let $A = [A_L\,\,A_R]$ where $A_L, A_R \in \K[x]^{n \times n/2}$
and let $\underline N, \overline N \in \K[x]^{n/2 \times n}$ be minimal
nullspace bases for, respectively, $A_L, A_R$.
This gives the first block-elimination step towards the diagonalization of $A$:
\begin{equation}\label{eq:recursion}
A = \begin{mat}{cc}A_L&A_R\end{mat}
\quad\to\quad
NA =\begin{mat}{c}\overline{N}\\\underline{N}\end{mat}
\!\!\begin{mat}{cc}A_L&A_R\end{mat}=\begin{mat}{cc}\overline{N}A_L&\\
& \underline{N}A_R\end{mat}.
\end{equation}
When $A$ is generic of degree $d$, 
it turns out that all the minimal indices of
both $\underline N$ and $\overline N$ are equal to $d$~\cite[Fact 1]{JeVi04}
and that $\overline{N}A_L$ and $\underline{N}A_R$ are $n/2 \times n/2$
polynomial matrices of degree exactly $2d$ on which we iterate. 

We show in~\cite{JeVi04} that the
property ``dimension $\times$ degree = $nd$'' generically carries from one iteration to the other: 
at step $i$, starting from $2^{i-1}$ blocks of dimensions $(n/2^{i-1}) \times (n/2^{i-1})$
and degree $2^{i-1} d$, 
we compute $2^{i-1}$ pairs $({\underline N}_i^{(j)},{\overline N}_i^{(j)})$ 
of minimal nullspace bases of dimensions $(n/2^i) \times (n/2^{i-1})$ 
and whose minimal indices are all equal to $2^{i-1} d$.
Let $(U,B)=(I_n,A)$ before the first step.
Step $i$ also requires to update the matrix transform as $U \gets \diag[ N_i^{(j)} ]_{j} \times U$
and the right hand side as $B \gets   \diag( N_i^{(j)} )_j  \times B$.
Because of the special block-structure of the polynomial matrices involved, 
it can be shown that
these updates reduce to solving $O(2^{2i})$ problems ${\sf MatMul}_{n/2^{i-1}, 2^{i-1} d}$.

Overall, the $\log n$ block-diagonalization steps thus reduce to ${\sf PartialNullSpace}_{n,d}^*$
and to 
\begin{equation} \label{eq:mstar}
\left\{\mbox{solve $O(2^{2i})$ problems ${\sf MatMul}_{n/2^i, 2^i d}$}\right\}_{0 \leq i < \log n}.
\end{equation}
By Lemma~\ref{lem:costpstar} and (\ref{eq:mstar}), we therefore obtain
a solution to ${\sf Inv}_{n,d}$ in $\so(n^3 d)$ operations in $\K$.

Since by Cramer's rule each entry of $A^{-1}$ has the form
$p/ (\det A)$ where $p \in \K[x]$ may have degree at large as $(n-1)d$,
the size of $A^{-1}$ is of the order of $n^3 d$. 
The above inversion algorithm, defined for $A$ generic and $n$ a power of two, 
is therefore nearly optimal.

\subsection{Determinant computation (${\sf Det}_{n,d}$)} \label{subsec:appli_det}

{\em Given $A \in \K[x]^{n \times n}$ of degree $d$, the problem is to compute $\det A \in \K[x]$.}\\

We assume here that $A$ is generic with $n$ is a power of two,
and we use the inversion algorithm of Section~\ref{subsec:appli_inv}.
It has been shown in~\cite{GJV03-2} that the diagonal
entries of the diagonal matrix $B$ in (\ref{eq:UA=B}) are constant
multiples of $\det A$. 
Since $\det A(0)$ is generically non-zero, we have
$$
\det A = \frac{\det A(0)}{b_{i,i}(0)} b_{i,i} \quad \mbox{for all $1 \leq i \leq n$}.
$$
The problem ${\sf Det}_{n,d}$ thus reduces essentially 
to computing the determinant of the constant matrix $A(0)$ 
and to the computation of, say, $b_{1,1}$.
It is well-known that over $\K$ computing the determinant reduces to matrix 
multiplication~\cite[Section 16.4]{BCS97}
(that is, ${\sf Det}_{n,0}$ reduces to ${\sf MatMul}_{n,0}$ using our notations).
Concerning $b_{1,1}$, we perform $\log n$ steps as for inversion but, 
since $b_{1,1}$ is the upper-left corner of $B$, 
we use instead of (\ref{eq:recursion}) the simpler step 
\begin{equation} \label{eq:recursion_det}
A = \begin{mat}{cc}A_L&A_R\end{mat}
\quad\to\quad {\overline N}A_L.
\end{equation}
As in (\ref{eq:recursion}), ${\overline N}$ is a minimal nullspace basis for $A_R$.
Step $i$ now consists in computing a single
minimal nullspace basis of dimensions $(n/2^i) \times (n/2^{i-1})$ 
and minimal indices $2^{i-1} d$,
and then in multiplying this basis with the left half of an $n/2^i$ by $n/2^i$ block of degree $2^{i-1}d$,
as in (\ref{eq:recursion_det}).
Hence, computing $b_{1,1}$ by performing these $\log n$ steps reduces to solving
${\sf PartialNullSpace}_{n,d}^*$ and ${\sf MatMul}_{n,d}^*$.
By Lemma~\ref{lem:costpstar}, this gives a solution to ${\sf Det}_{n,d}$ in $\so(n^\omega d)$
operations in $\K$.

Notice that when $A$ is not generic or when $n$ is not a power of two, 
a Las Vegas $\so(n^\omega d)$ solution to ${\sf Det}_{n,d}$ can be obtained
using the Smith normal form algorithm in~\cite{Sto03}.

\subsection{Row reduction (${\sf RowRed}_{n,d}$)} \label{subsec:appli_rowred}

{\em Given $A \in \K[x]^{n \times n}$ of degree $d$, the problem is to
compute $R \in \K[x]^{n \times n}$ that is row-reduced and unimodularly left equivalent to $A$.}\\\

We assume here that $A(0)$ is non-singular.
Recall from Section~\ref{subsec:mab} and~\cite[\S6.3.2]{Kai80} that 
$R = A$ is a row-reduced form of $A$ when $R$ is row-reduced and 
$R=UA$ for some unimodular polynomial matrix $U$.
The solution in~\cite{GJV03-2} works by expansion/reconstruction of the 
matrix fraction $H$ as in Theorem~\ref{theo:expproper} with $h=(n-1)d+1$.

First, we expand $H$ up to order $2d+1$.
This is done by solving ${\sf MatFracExp}_{n,2d+1}$ once, 
taking $B = I_n$ and $h = (n-1)d+1 = O(nd)$.
From Theorem~\ref{theo:expproper} we know that 
$H$  is a strictly proper matrix fraction 
which admits left and right factorizations $A^{-1} (AH)$ and $(HA) A^{-1}$.
Strict properness further implies that the degrees of both $AH$ and $HA$
must be less than the degree of $A$~\cite[Lemma 6.3-10]{Kai80}, and are thus bounded by $d$ as well.
Therefore, these left and right factorizations of $H$ are factorizations of degree $d$
and, using Theorem~\ref{theo:recon}, we can reconstruct $H$ from its
expansion up to order $2d+1$ as $H = R^{-1}S$.
This reconstruction corresponds to solving problem ${\sf MatFracRec}_{n,d}$ once.
On one hand, we know by Theorem~\ref{theo:recon} that $R$ is row-reduced.
On the other hand, $A^{-1} (AH)$ and $R^{-1} S$ are coprime factorizations
of the same fraction, which implies that there exists a unimodular $U$
such that $UA = R$~\cite[Theorem 6.5-4]{Kai80}.
It follows that $R$ is indeed a row-reduced form of $A$.
By Lemma~\ref{lem:costpstar}, 
this reduction to  ${\sf MatFracExp}_{n,2d+1}$ and ${\sf MatFracRec}_{n,d}$
gives a  solution to ${\sf RowRed}_{n,d}$ in $\so(n^\omega d)$ operations in $\K$.

\subsection{Small nullspace computation (${\sf Nullspace}_{n,d}$)}  \label{subsec:appli_nullspace}

{\em Given $A \in \K[x]^{n \times n}$ of degree $d$,  
the problem is to compute the rank $r$ of $A$, and 
$N \in \K[x]^{(n-r) \times n}$ of rank $n-r$
such that $N A = 0$.}\\\

As already seen, a solution in the restrictive ({\em e.g.} generic) case when all minimal
vectors have degrees in $O(d)$ is provided by a solution 
to ${\sf PartialNullSpace}_{n,d}$. In the general case 
the row degrees in a nullspace basis of $A$ may be unbalanced, they range 
between $0$ and $nd$~\cite[Theorem~3.3]{StVi05-2}.
Previously known methods, whose cost is essentially driven by the highest
Kronecker index, do not seem to allow the target complexity estimate
$\so(n^{\omega}d)$ (see for instance~\cite[Section~2]{StVi05-2}).

Our solution in~\cite{StVi05-2} first reduces the general nullspace 
problem to
the full column rank case via randomization. This consists in
evaluating the rank $r$ of $A$ at a random $x=x_0$, then in compressing 
$A$ to a full column rank matrix. We also derive a particular
strategy when $n \gg r$.
Consequently, for a simplified explanation here, we now assume 
that $A$ has full column rank $n$ and
dimensions $(n+m) \times n$ with $m=O(n)$.   

The algorithm then works in $i$ steps with $1 \leq i
\leq \log n$. At step $i$ we compute a set of about $m/2^i$ nullspace
vectors of degrees less that $\delta = 2 ^i d$. These vectors are
obtained from 
$\log n$ solutions to ${\sf PartialNullSpace}_{m,\delta}$
for nullspace vectors of bounded degree $\delta=2 ^i d$,
and involving matrices of decreasing dimensions $n+m /2^i$.
Hence we essentially have a reduction to ${\sf PartialNullSpace}^*_{m,\delta}$.
We may point out that the proof of Theorem~\ref{theo:partialnullspace}
for the cost of the partial nullspace 
itself relies on solutions to 
${\sf MatFracExp}_{m,\delta}$, and 
${\sf MatFracRec}_{m,\delta}$. Nullspace vectors are 
computed using a matrix fraction expansion\,/reconstruction 
scheme.

The appropriate instances for 
${\sf PartialNullSpace}_{m/2^i,2^id}$, $1 \leq i
\leq \log n$, are built as submatrices of the input matrix $A$. 
Our choices for these submatrices ensure the linear independency of 
the successive computed sets of nullspace vectors.  
The algorithm hence outputs a union of a logarithmic number of sets
of 
linearly independent nullspace vectors. 
Each set, corresponding to an instance of ${\sf
  PartialNullSpace}_{m/2^i,2^id}$,
is a family of minimal vectors for a submatrix of $A$. The
minimality is not preserved in general with respect to $A$, however
we prove that small degree vectors are
obtained~\cite[Proposition~7.1]{StVi05-2}.

This reduction of ${\sf NullSpace}_{n,d}$
to ${\sf PartialNullSpace}^*_{m,\delta}$ and 
to ${\sf MatMul}^*_{n,d}$
for additional matrix multiplications
establishes that 
a solution matrix $N$ such that $NA=0$ can be computed in
$\so(n^{\omega}d)$ operations in $\K$ by a randomized Las Vegas (certified) algorithm.

\subsection{Factorization (${\sf Factor}_{n,d}$)} \label{subsec:appli_factor} 

{\em Given a right factorization $BA ^{-1}$ of degree $d$ of $H \in
\K(x)^{n \times n}$, the problem is to compute polynomial matrices 
$U$ and $V$ such that $V ^{-1}U=H$.}\\\ 

Corollary~\ref{cor:trivial_recon},  together with the expansion of $H=B A
^{-1}$,  provides
a solution to 
${\sf FracMatRec}_{n,d}$ if $H$ admits factorizations of degree $d$ on both sides. 
The solution of the general case, we mean for an arbitrary left side factorization,  
induces several difficulties for dealing with unbalanced row degrees.
These difficulties are bypassed using the techniques of Section~\ref{subsec:appli_nullspace}.

By considering the polynomial matrix $[-A ^T~~B^T]$ 
and solving ${\sf Nullspace_{2n,d}}$ we get $U$ and $V$ such that
$$
[U~~V]\left[\begin{array}{c} -A \\ B \end{array}\right]=0.
$$
Arguments similar to those used in the proof of
Theorem~\ref{theo:recon}
lead to the fact that $V$ is non-singular. Hence a solution $V
^{-1}U$ to the factorization problem is computed in $\so
(n^{\omega}d)$ operations in $\K$.
Note that since a solution to ${\sf Nullspace_{2n,d}}$
may not be minimal, the factorization $V ^{-1}U$ may not be coprime.

\bibliographystyle{plain}
{\small 
\bibliography{approx}
}

\end{document}